\documentclass[12pt,journal]{IEEEtran}
\onecolumn
\usepackage{epsfig}
\usepackage{subfigure}

\begin{document}

\title{Distinguishing Computer-generated Graphics from Natural Images Based on Sensor Pattern Noise and Deep Learning}

\author{Ye~Yao,~%\IEEEmembers,hip{Member,~IEEE,}
		Weitong Hu, Wei Zhang, Ting Wu and Yun-Qing Shi
\thanks{This work was supported by the Zhejiang Provincial Natural Science Foundation of China (No.LY14F020044), the Key Research and Development Program of Zhejiang Province (No. 2017C01062), the Public Technology Application Research Project of ZheJiang Province (No. 2017C33146), the Humanities and Social Sciences Foundation of Ministry of Education of China (No. 17YJC870021), and the National Natural Science Foundation of China (No. 61772165) (\emph{Corresponding author: magherozhw@hdu.edu.cn (W.Z), wuting@hdu.edu.cn (T.W)})}
\thanks{Ye Yao is with the School of CyberSpace, Hangzhou Dianzi University, 
Hangzhou, 310018, China. (e-mail: yyaoprivate@gmail.com).}% <-this % stops a space
\thanks{Weitong Hu is with the School of CyberSpace, Hangzhou Dianzi University, 
Hangzhou, 310018, China. (e-mail: hwt@hdu.edu.cn).}
\thanks{Wei Zhang is with the School of Computer Science and Technology, Hangzhou Dianzi University, Hangzhou, 310018, China. (e-mail: magherozhw@hdu.edu.cn).}
\thanks{Ting Wu is with the School of CyberSpace, Hangzhou Dianzi University, 
Hangzhou, 310018, China. (e-mail: wuting@hdu.edu.cn).}
\thanks{Yun-Qing Shi is with the Department of Electrical and Computer Engineering, New Jersey Institute of Technology, Newark, NJ 07102 USA (e-mail: shi@njit.edu).}% <-this % stops a space
}

%\markboth{IEEE SIGNAL PROCESSING LETTERS,~VOL.~XX,~NO.~X, XXX~2018}%
%{YAO \MakeLowercase{\textit{et al.}}: Deep Learning for Object Forgery Detection in Advanced Video}

\maketitle

\begin{abstract}
Computer-generated graphics are images generated by computer software. The rapid development of computer graphics technologies has made it easier to generate a photorealistic computer graphics, and these graphics are quite difficult to distinguish from natural images by our naked eyes. In this paper, we propose a method based on sensor pattern noise and deep learning to distinguish computer-generated graphics (CG) from natural images (NI). Before being fed into our convolutional neural network (CNN)-based model, these images---including the CG and NI---are clipped into image patches. Furthermore, several high-pass filters (HPF) are used to remove low-frequency signal, which represents the image content. These filters are also used to enhance the residual signal as well as sensor pattern noise introduced by the digital camera device. Different from the traditional methods of distinguishing CG from NI, the proposed method utilizes a five-layer CNN to classify the input image patches. Based on the classification results of the image patches, we deploy a majority vote scheme to obtain the classification results for the full-size images. The experiments have demonstrated that: 1) the proposed method with three high-pass filters can achieve better results than that with only one high-pass filter or no high-pass filter. 2) the proposed method with three high-pass filters achieves 100\% accuracy, although the natural images undergo a JPEG compression with a quality factor of 75.
\end{abstract}

% Note that keywords are not normally used for peer review papers.

\section{Introduction}

\begin{figure}%[htbp]
	\begin{minipage}[t]{0.32\textwidth}
	\centering
	\includegraphics[width=\textwidth]{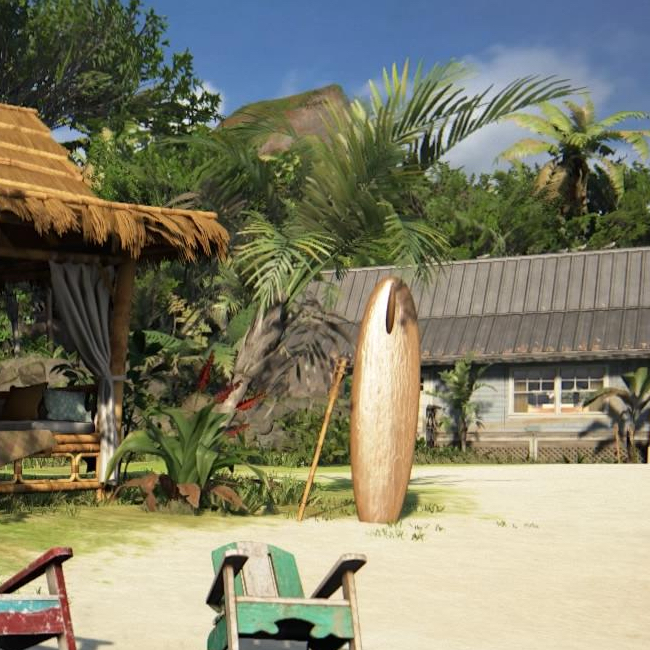}
	\end{minipage}
	\begin{minipage}[t]{0.32\textwidth}
	\centering
	\includegraphics[width=\textwidth]{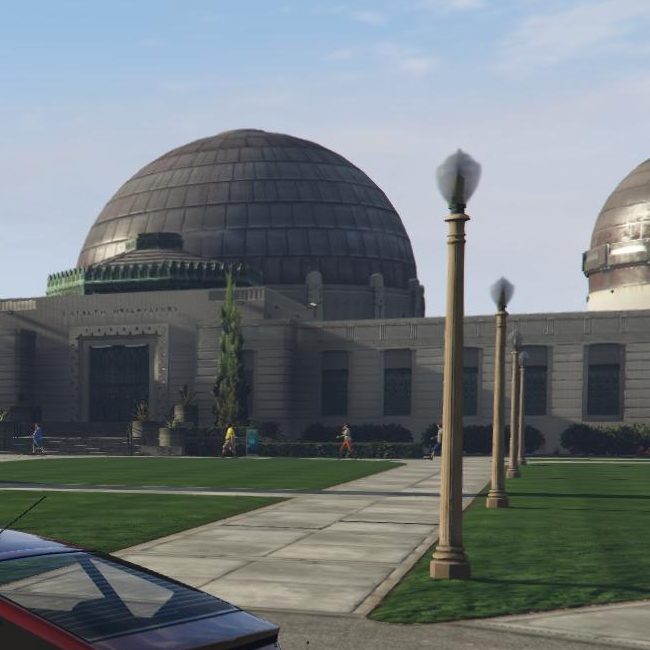}
	\end{minipage}
	\begin{minipage}[t]{0.32\textwidth}
	\centering
	\includegraphics[width=\textwidth]{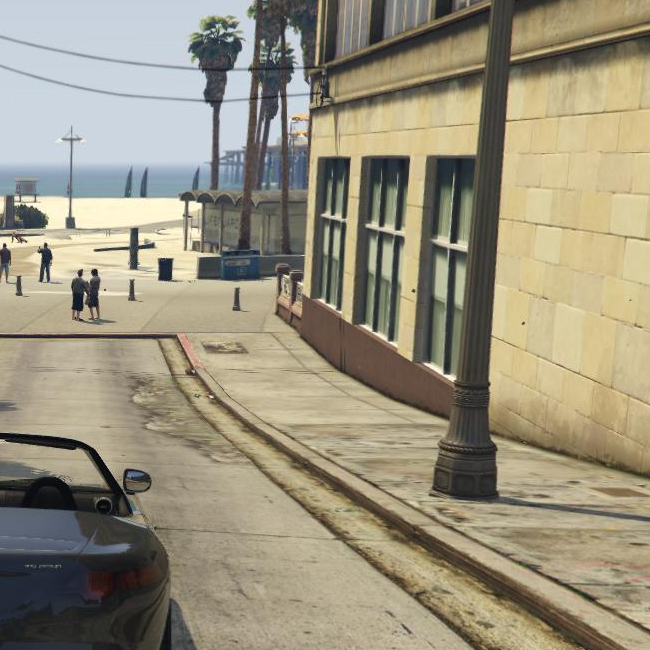}
	\end{minipage}
	\vfill
	\begin{minipage}[t]{0.32\textwidth}
	\centering
	\includegraphics[width=\textwidth]{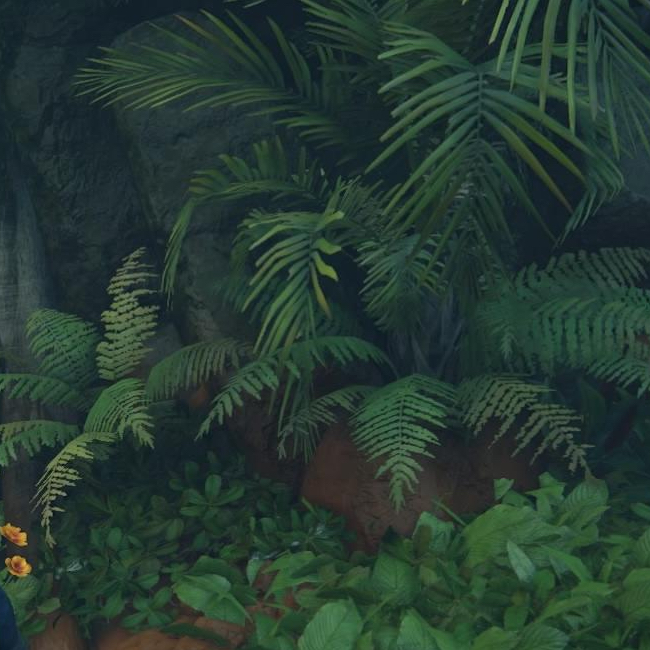}
	\end{minipage}
	\begin{minipage}[t]{0.32\textwidth}
	\centering
	\includegraphics[width=\textwidth]{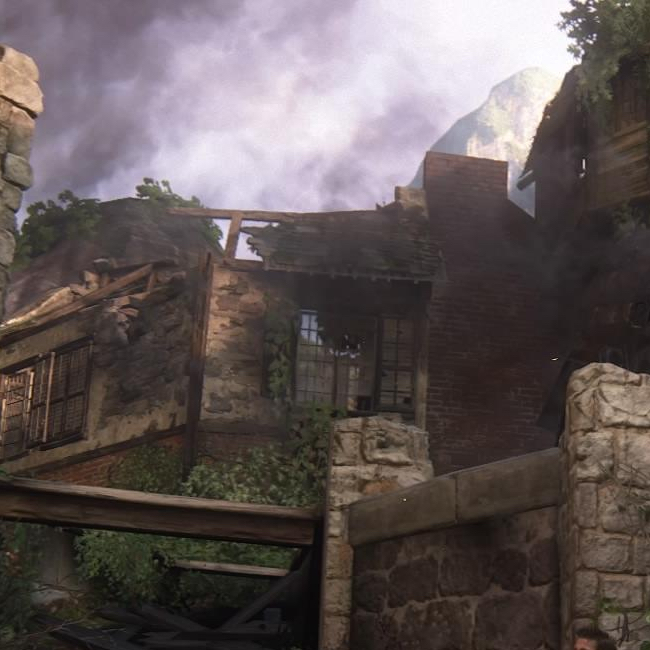}
	\end{minipage}
	\begin{minipage}[t]{0.32\textwidth}
	\centering
	\includegraphics[width=\textwidth]{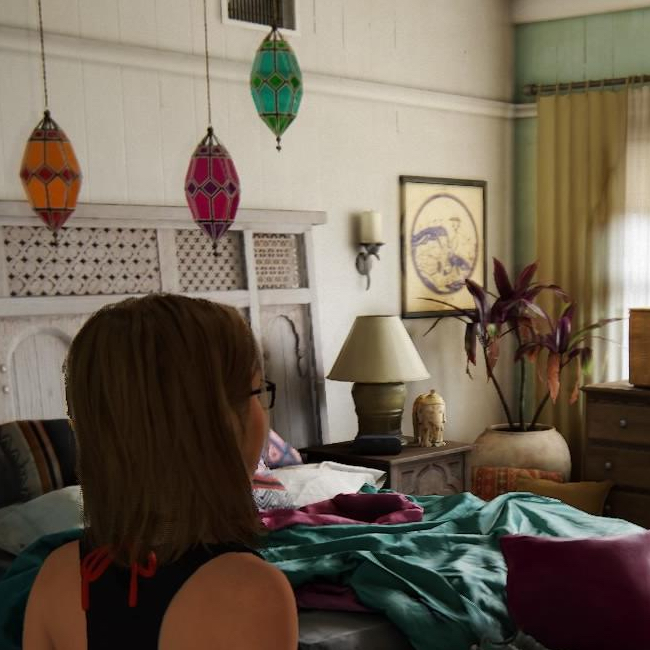}
	\end{minipage}	
	\caption{Some examples of computer graphics (CG) from dataset in~\cite{Rahmouni:2017}.}
	\label{fig:cg}
\end{figure}

\IEEEPARstart{C}{omputer}-generated graphics (CG) are images generated by computer software such as 3D Max, Maya, and Cinema 4D. In recent years, with the aid of computer software, it is easier to generate photorealistic computer graphics (PRCG), which are quite difficult to distinguish from natural images (NI) by the naked eye. Some examples of computer graphics are shown in Figure~\ref{fig:cg}. Although these rendering software suites help us to create images and animation conveniently, it could also bring serious security issues to the public if PRCG were used in fields such as justice and journalism~\cite{Rocha:2011}. Therefore, as an essential topic in the domain of digital image forensics~\cite{Stamm:2013}, distinguishing CG from NI has attracted increasing attention in the past decade.

Several algorithms have recently been proposed to distinguish computer-generated graphics from natural images. Xiaofen Wang~{et al.}~\cite{Wang:2014} present a customized statistical model based on the homomorphic filter and use support vector machines (SVMs) as a classifier to discriminate photorealistic computer graphics (PRCG) from natural images. Zhaohong Li~{et al.}~\cite{Li:2015} present a multiresolution approach to distinguish CG from NI based on local binary patterns (LBPs) features and SVM classifier. Jinwei Wang~{et al.}~\cite{Wang:2016} present a classification method based on the first four statistical features extracted from the quaternion wavelet transform (QWT) domain. Fei Peng~{et al.}~\cite{Peng:2017} present a method to extract 24 dimensions of features based on multi-fractal and regression analysis for the discrimination of computer-generated graphics and natural images. However, all of these methods have depended on handcrafted features from computer-generated graphics and natural images, and also depend on SVM as the classifier.

Deep learning has been used in many new fields and has achieved great success in recent years. Deep neural networks such as the convolutional neural network (CNN) have the capacity to automatically obtain high-dimensional features and reduce its dimensionality efficiently \cite{Yao:2018}. Some researchers have begun to utilize deep learning to solve problems in the domain of image forensics, such as image manipulation detection~\cite{Bayar:2016}, camera model identification~\cite{Bondi:2017,Tuama:2016}, steganalysis~\cite{Xu:2016}~\cite{J.:2017}, image copy-move forgery detection~\cite{Rao:2016}, and so on. 

In this paper, we propose a method based on sensor pattern noise and deep learning to distinguish computer-generated graphics (CG) from natural images (NI). The main contributions are summarized as follows: 1) Different from the traditional methods of distinguishing CG from NI, the proposed approach utilizes a five-layer convolutional neural network (CNN) to make a classification for the input images. 2) Before being fed into the CNN-based model, these images---including the CG and NI---are clipped to image patches. 3) Several high-pass filters (HPFs) are used to remove low-frequency signal, which represents the image contents. These filters are also used to enhance the residual signal as well as sensor pattern noise introduced by the digital camera device. 4) The experimental results have shown that the proposed method with three high-pass filters can achieve 100\% accuracy, although the natural images undergo a JPEG compression with a quality factor of 75.

%%%%%%%%%%%%%%%%%%%%%%%%%%%%%%%%%%%%%%%%%% section 2
\section{Related Works}
In this paper, we propose a method of distinguishing computer-generated graphics from natural images based on sensor pattern noise and deep learning. There are several studies related to deep learning as well as sensor pattern noise used for forensics.

\subsection{Methods Based on Deep Learning}
Gando~{et al.}~\cite{Gando:2016} presented a deep learning method based on a fine-tuned deep convolutional neural network. This method can automatically distinguish illustrations from photographs and achieve 96.8\% accuracy. It outperforms other models, including custom CNN-based models trained from scratch and  traditional models using handcrafted features.

Rahmouni~{et al.}~\cite{Rahmouni:2017} presented a custom pooling layer to extract statistical features and a CNN framework to distinguish computer-generated graphics from real photographic images. A weighted voting scheme was used to aggregate the local estimates of class probabilities and predict the label of the whole picture. The best accuracy in ~\cite{Rahmouni:2017} is 93.2\%, obtained by the proposed Stats-2L model.

\subsection{Sensor Pattern Noise Used for Forensics}
Different digital cameras introduce different noise to their output digital images. The main noise sources are due to the imperfection of CCD or CMOS sensors. It has been named as sensor pattern noise (SPN) and is used as a fingerprint to characterize an individual camera. In particular, SPN has been used in image forgery detection~\cite{Pandey:2016} and source camera identification~\cite{Orozco:2016}.

Villalba~{et al.}\cite{Villalba:2016} presented a method for video source acquisition identification based on sensor noise extraction from video key frames. Photo response non-uniformity (PRNU) is the primary part of the sensor pattern noise in an image. In~\cite{Villalba:2016}, the PRNU is used to calculate the sensor pattern noise and characterize the fingerprints into feature vectors. Then, the feature vectors are extracted from the video key frames and trained by a SVM-based classifier.

%%%%%%%%%%%%%%%%%%%%%%%%%%%%%%%%%%%%%%%%%% section 3

\section{Proposed Method}
The proposed method consists of two primary steps: image preprocessing and CNN-based model training. In the first step, the input images---including the computer-generated graphics and the natural images---are clipped to image patches, then three types of high-pass filter (HPF) are applied to the image patches. These filtered image patches constitute the positive and negative training samples. In the second step, the filtered image patches are fed to the proposed CNN-based model for training. The proposed CNN-based model is a five-layer CNN. In this section, we introduce these steps of our method in detail.

\subsection{Image Preprocessing}
\subsubsection{Clipped to Image Patches}
The natural images taken by cameras and the computer graphics generated by software often have a large resolution. Due to hardware memory limitations, we need to clip these full-size images into smaller image patches before they are fed into our neural network for training. This is also a data augmentation strategy in deep learning approaches to computer vision~\cite{Yao:2018}. Data augmentation~\cite{Krizhevsky:2012} helps to increase the amount of training samples used for deep learning training and improve the generalization capability of the trained model. Therefore, we propose to clip all of the full-size images to image patches. The resolution of each image patch is 650$\times$650. We chose this size as a trade-off between processing time and computational limitations.

Both the computer-generated graphics and the natural images are clipped into image patches. All of the clipping is label-preserving operations. That is to say, we prepare the positive samples by  drawing image patches from the full-size natural images. In a similar way, we get negative samples from the full-size computer-generated graphics. However, natural images taken by cameras usually have a larger resolution than computer-generated graphics. If we want the amount of negative samples and the amount of positive samples to be approximately equivalent, we need to clip more image patches in each computer-generated graphic than we do from natural images. In light of this, we set the stride size for natural images to the width of the image patches (i.e., 650). After analyzing the amount of the image patches, we set the stride size for computer-generated graphics to a smaller value (i.e., 65).

\begin{figure}
\subfigure[SQUARE5$\times$5]{
	\begin{minipage}[t]{0.32\textwidth}
	\centering
	\includegraphics[width=0.8\textwidth]{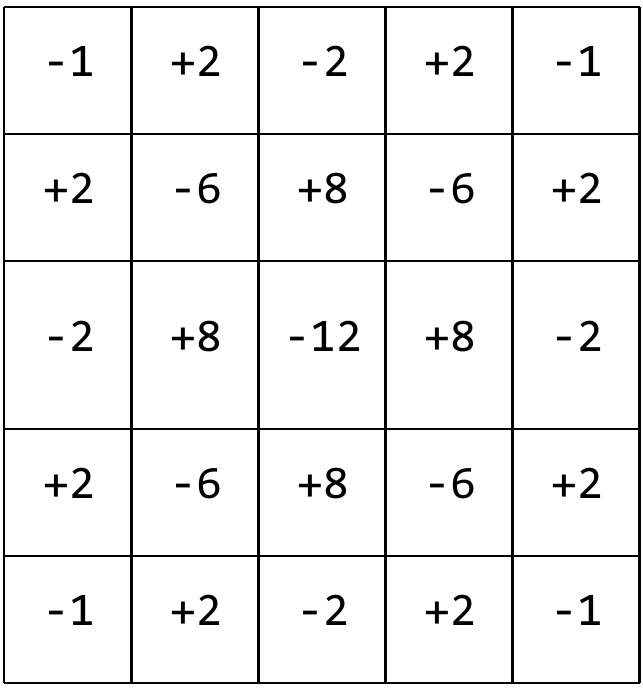}
	\end{minipage}	
}
\subfigure[EDGE3$\times$3]{
	\begin{minipage}[t]{0.32\textwidth}
	\centering
	\includegraphics[width=0.8\textwidth]{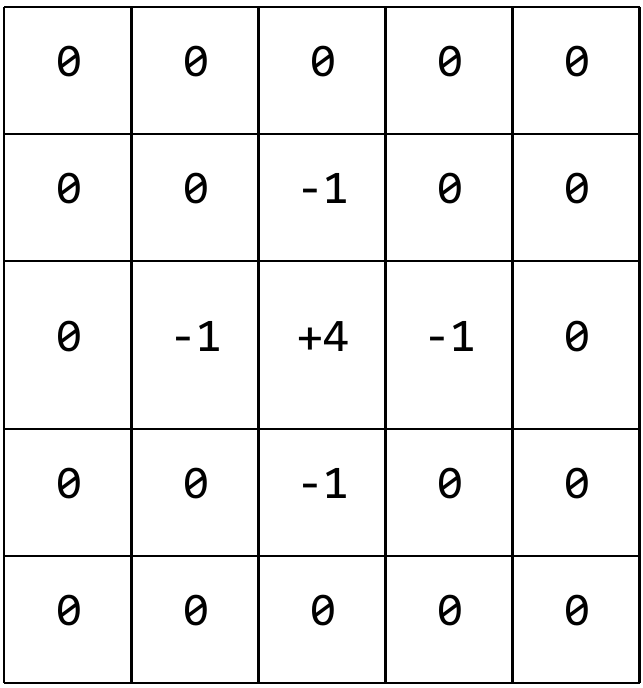}
	\end{minipage}
}
\subfigure[SQUARE3$\times$3]{	
	\begin{minipage}[t]{0.32\textwidth}
	\centering
	\includegraphics[width=0.8\textwidth]{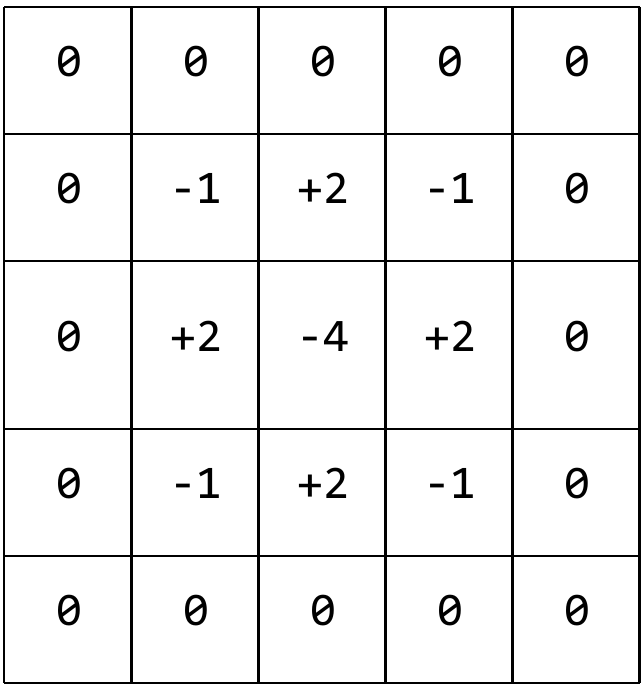}
	\end{minipage}
}
	\caption{Three types of high-pass filter (HPF) used in the proposed method.}
	\label{fig:hpf}
\end{figure}

\subsubsection{Filtered with High-Pass Filter}
Since the natural images and the computer-generated graphics are created from different pipelines, there should exist some distinct differences between them. As we all know, sensor pattern noise (SPN) has been used to identify the source camera of a natural image, and has obtained excellent performance~\cite{Tuama:2016,Bondi:2017}~\cite{Orozco:2016}. However, there is no sensor pattern noise in computer-generated graphics. Based on this idea, we propose our method to discriminate  computer-generated graphics from natural images.

Fridrich~{et al.}~\cite{Fridrich:2012} designed several high-pass filters for the steganalysis of digital images. As it is mentioned, these filters have the ability to obtain the noise residuals and suppress the value of the low-frequency component, which represents the image content. Qian~{et al.}\cite{Qian:2015} proposed a customized convolutional neural network for steganalysis. This customized deep learning approach starts with a predefined high-pass filter. This predefined HPF was proposed as a noise residual model of SQUARE5$\times$5 in~\cite{Fridrich:2012}. Furthermore, this noise residual model has been applied to deep learning-based camera model identification~\cite{Tuama:2016} as well as to deep learning-based video forgery detection~\cite{Yao:2018}, and has obtained perfect performance.

In this paper, we utilize several high-pass filters in our method to obtain the sensor noise residuals and reduce the impact of the image content. These predefined high-pass filters are employed to make a convolution operation with the image patches. Furthermore, in order to reduce the computational complexity, the image patches are first converted to grayscale. The predefined high-pass filters are applied to the grayscale image patches, then the noise residuals of the image patches are piped into the proposed convolutional neural network.

The proposed high-pass filters are shown in Figure~\ref{fig:hpf}. There are three types of high-pass filter used in our method. The SQUARE5$\times$5 and SQUARE3$\times$3 were proposed as noise residuals model in~\cite{Fridrich:2012}. The EDGE3$\times$3 was designed by us according to the different structure of all the other filters in~\cite{Fridrich:2012}. In order to let the three filters have the same size, the elements in the bounding boxes of the SQUARE3$\times$3 and the EDGE3$\times$3 are set to zero.

\subsection{CNN-Based Model Training}
\begin{figure}%[htbp]
\centering
\includegraphics[width=8cm]{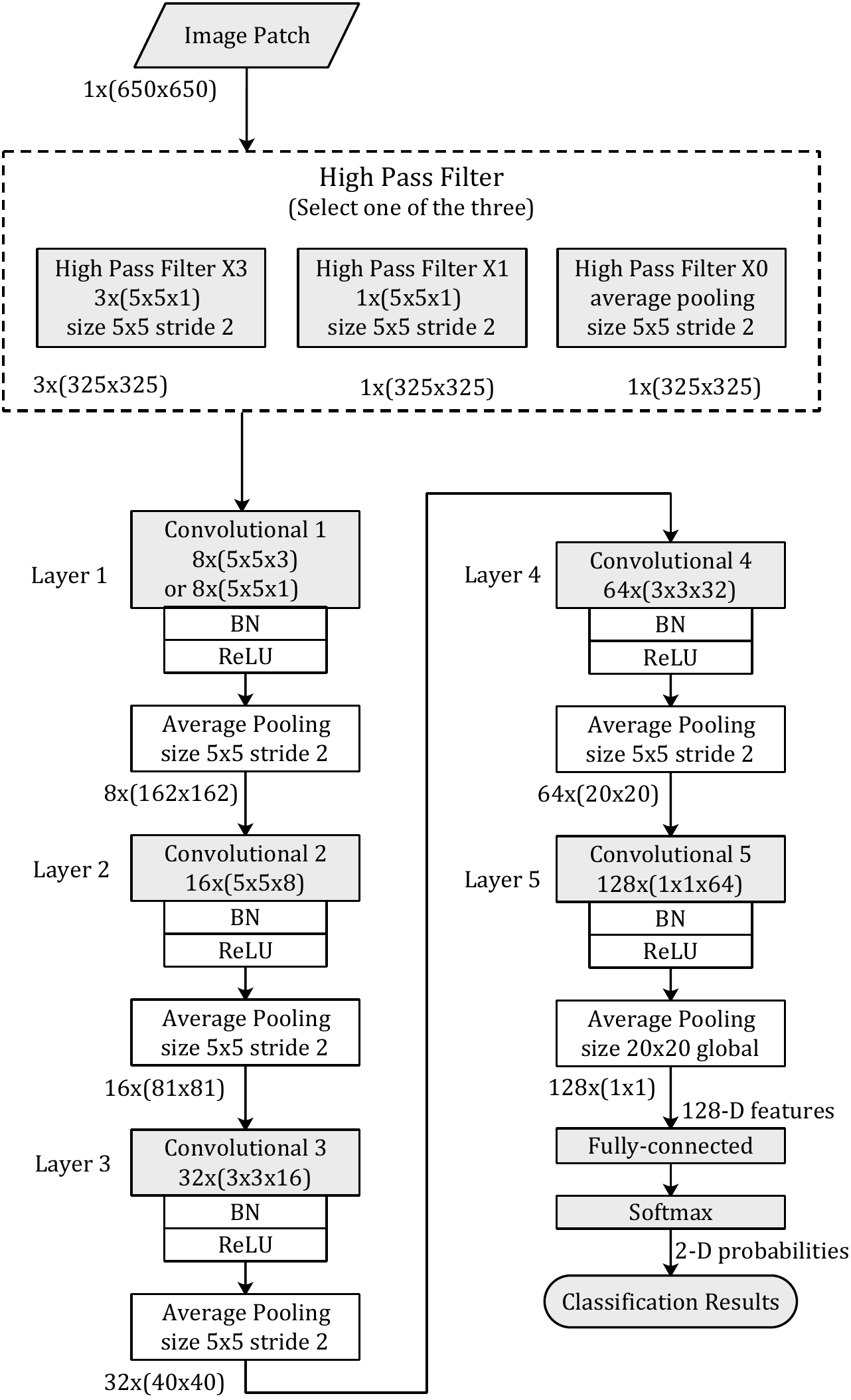}
\caption{The proposed convolutional neural network architecture. Names and parameters of each layer are displayed in the boxes. Kernel sizes in each convolution layer are shown as $number\_of\_kernels \times (width\times height\times number\_of\_input)$. Sizes of feature maps between consecutive layers are shown as $number\_of\_feature\_maps\times (width\times height)$. Padding is used in each convolutional layer to keep the shape of image patches. BN: batch normalization; ReLU: rectified linear units.}
\label{fig:cnn}
\end{figure}

The proposed convolutional neural network architecture is illustrated in Figure~\ref{fig:cnn}. The image patches of the input for the proposed neural network are image blocks clipped from the full-size natural images or computer-generated graphics with a resolution of 1$\times$(650$\times$650), where 1 represents the channel number of gray-scale, 650 represents the size of width and height.

There is a high-pass filter layer at the top of the proposed CNN-based model. This filter layer consists of three combinations of high-pass filters. We need to select one type of the three combinations for the deep learning training. The combination of the \emph{High\_Pass\_Filter$\times$3} consists of all three proposed filters; i.e., the SQUARE5$\times$5, the EDGE3$\times$3, and the SQUARE3$\times$3. The combination of the \emph{High\_Pass\_Filter$\times$1} only contains the SQUARE5$\times$5 filter. The combination of the \emph{High\_Pass\_Filter$\times$0} utilizes an average pooling layer instead of the high-pass filter layer. According to the combination used by the proposed method, the number of  feature maps outputted by the filter layer is different. If the \emph{High\_Pass\_Filter$\times$3} is used, there will be three feature maps with size 325$\times$325 outputted by the high-pass filter layer. Otherwise, there will only be one feature map of size 325$\times$325 outputted by the high-pass filter layer.

The proposed CNN architecture consists of five convolutional layers. Each convolutional layer is followed by a batch normalization (BN)~\cite{Ioffe:2015} layer, a rectified linear units (ReLU)~\cite{Nair:2010} layer, and an average pooling layer. At the bottom of the proposed model, a fully-connected layer and a softmax layer are utilized to transform the 128 dimensional feature vectors to classification probabilities of the image patches.

The kernel sizes of the convolution layers in the proposed CNN-based model are 5$\times$5, 5$\times$5, 3$\times$3, 3$\times$3, and 1$\times$1, respectively. The amounts of the feature maps outputted by each layer are 8, 16, 32, 64, and 128, respectively, and the size of feature maps are 325$\times$325, 162$\times$162, 81$\times$81, 40$\times$40, and 20$\times$20, respectively. The kernel size of the average pooling in each layer is 5$\times$5 and the stride size is 2. Note that the last average pooling layer has a global kernel size of 20$\times$20.

%%%%%%%%%%%%%%%%%%%%%%%%%%%%%%%%%%%%%%%%%%%% section 4
\section{Experiments}
\subsection{Dataset}
We compared our deep learning approach with the state-of-the-art methods in~\cite{Rahmouni:2017}. The dataset used in this paper is the same as the dataset in~\cite{Rahmouni:2017}. It consists of 1800 computer-generated graphics and 1800 natural images. The computer-generated graphics were downloaded from the Level-Design Reference Database~\cite{leveldb:2017}, which contains more than 60,000 screenshots of photo-realistic video games. The game information was removed by cropping the images to a resolution of 1280$\times$650. The preprocessed images can be downloaded from the link on Github~\cite{githubcode:2017}. Some computer graphics samples are shown in Figure~\ref{fig:cg}. The natural images are taken from the RAISE dataset~\cite{Dang-Nguyen:2015}. The resolution of these natural images ranges from 3008$\times$2000 to 4900$\times$3200. All of these natural images were downloaded in RAW format and converted to JPEG with a quality factor of 95.

In our experiment, 900 natural images and 900 computer-generated graphics were randomly selected from the dataset for training, 800 natural images and 800 computer-generated graphics were set aside for testing, and 100 natural images and 100 computer-generated graphics for validation. Then, all of these full-size images were clipped to image patches with size 650$\times$650. The number of image patches we obtained for training was about 44,000.

\subsection{Experiment Setup}
We implemented the proposed convolution neural network based on the Caffe framework~\cite{Jia:2014}. All of the experiments were conducted on a GeForce GTX 1080ti GPU. The stochastic gradient descent algorithm was used to optimize the proposed CNN-based model. The initial learning rate was set to 0.001. The learning rate update policy was set to $inv$ with the $gamma$ value of 0.0001 and the $power$ value of 0.75. The parameters of $momentum$ and $weight\_decay$ were set to 0.9 and 0.0005, respectively. The batch size of training was set to 64. Namely, 64 image patches were fed to the CNN-based model for each iteration. After 80 epochs of iteration, the trained CNN-based model was obtained for testing.

In order to get the performance of the proposed CNN-based model, we applied the trained model to the testing dataset. All of the full-size images in the testing dataset needed to be preprocessed. The preprocessing for the testing images was similar to the preprocessing of the images in the training. After preprocessing, the testing images were clipped into image patches. Then, these image patches were fed to the trained CNN-based model, and the prediction results for the image patches were obtained. Based on the prediction results of the image patches, we deployed a majority vote scheme to obtain the classification results for the full-size images.

\begin{figure}%[htbp]
\centering
\includegraphics[width=12cm]{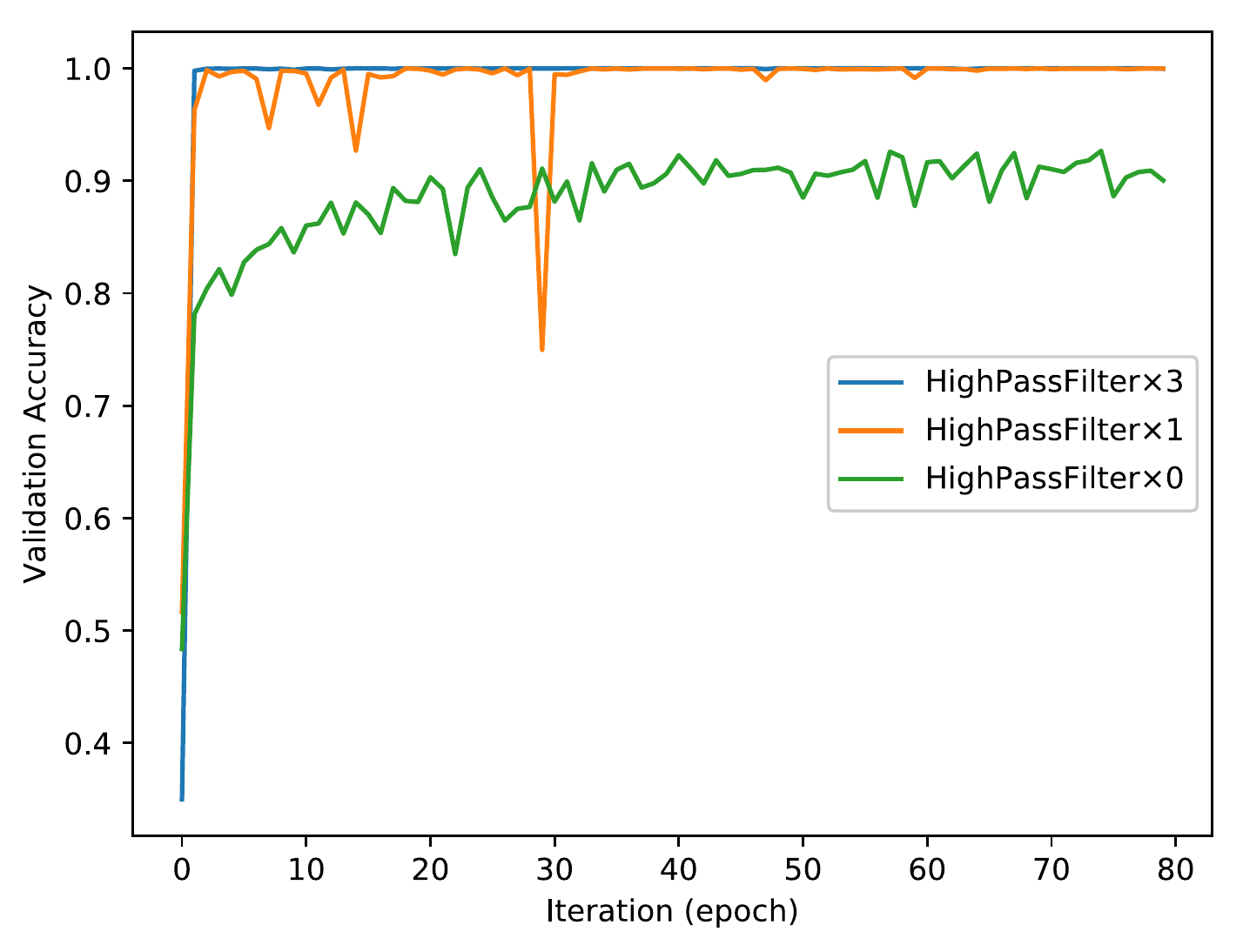}
\caption{Validation performance of the proposed method.}
\label{fig:hpf-acc}
\end{figure}

\begin{figure}%[htbp]
\centering
\includegraphics[width=12cm]{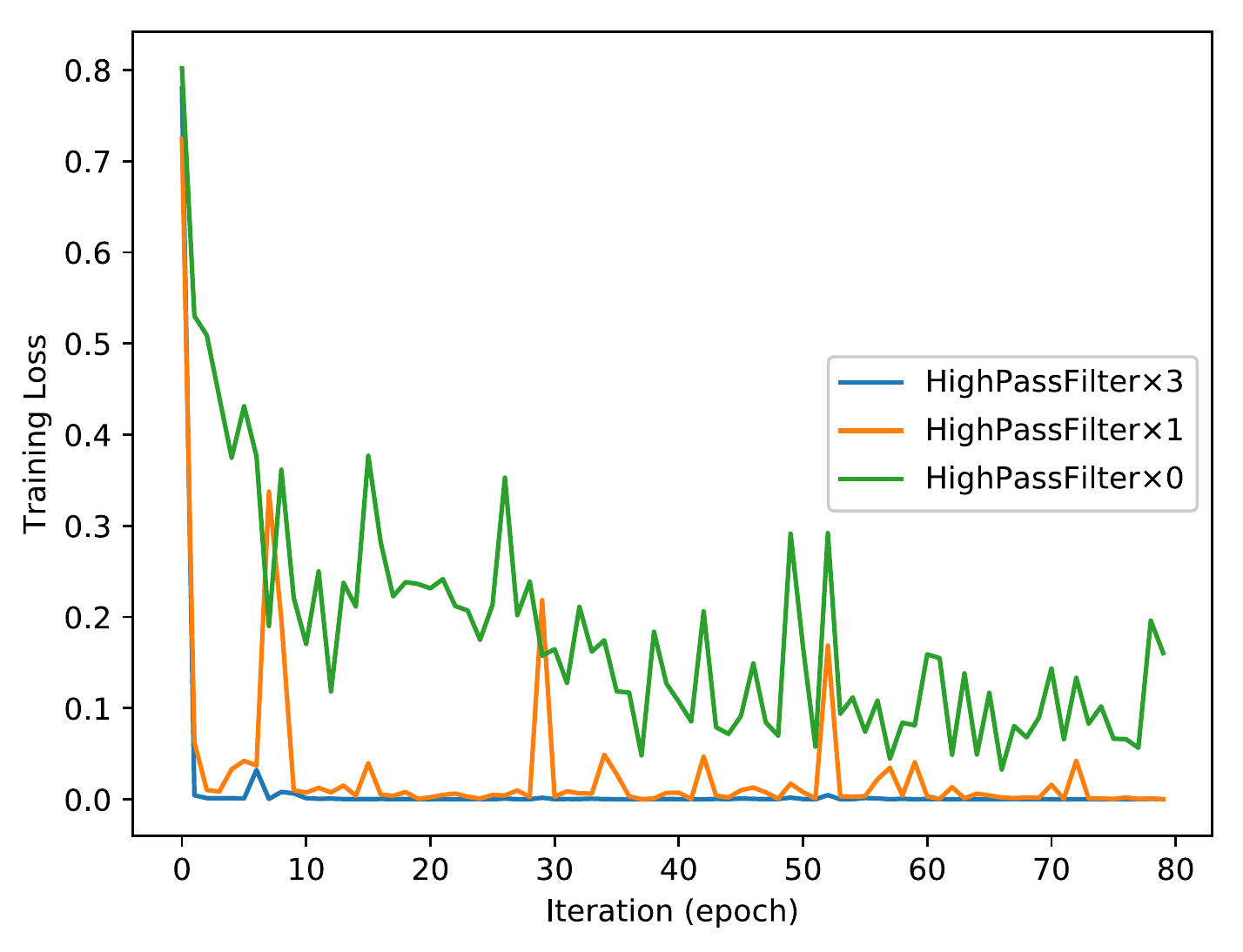}
\caption{Training loss of the proposed method.}
\label{fig:hpf-loss}
\end{figure}

\subsection{Experimental Results}
\subsubsection{Different Numbers of High-Pass Filters}
As shown in Figure~\ref{fig:cnn}, the proposed convolution neural network has three combinations for the high-pass filter layer. Each of the combinations has different numbers of high-pass filters. We trained all of these combinations for 80 epoch iterations and obtained two trained models for each of the combinations. In other words, we obtained a model of 50 epochs and a model of 80 epochs for the combination of ~\emph{High\_Pass\_Filter$\times$3} after training the proposed network for 80 epoch iterations. We also obtained the same number of models for the other two combinations. Figure~\ref{fig:hpf-acc} and Figure~\ref{fig:hpf-loss} show the evolutions of training loss and validation accuracy in the procedure of iteration. The validation accuracy is shown in Figure~\ref{fig:hpf-acc}, and the training loss is shown in Figure~\ref{fig:hpf-loss}. It is observed that the proposed method with ~\emph{High\_Pass\_Filter$\times$3} converges much faster than the others and achieves much higher prediction accuracy. 

To evaluate the classification performance of the proposed method with different numbers of high-pass filters, we tested these models obtained in the training procedure on the testing dataset. The classification accuracy is shown in Table~\ref{table-hpf}. Note that the size of the image patches in the method of Rahmouni~{et al.} in~\cite{Rahmouni:2017} is 100$\times$100. In our experiments, we set the size of the image patches to 650$\times$650 to meet the requirement of our neural network architecture. A majority vote scheme was applied to the testing results of the image patches to obtain the classification results for the full-size images. 

Compared with the state-of-the-art method of Rahmouni~{et al.} in~\cite{Rahmouni:2017}, our method with the high-pass filter obtained better performance. Furthermore, the proposed method with ~\emph{High\_Pass\_Filter$\times$3} outperformed the others and obtained the best performance. The classification accuracy for the full-size images could achieve 100\%. These experimental results demonstrate the effectiveness of the high-pass filter in the preprocessing procedure for our proposed deep learning approach.

\renewcommand\arraystretch{1.5}
\begin{table}[htbp]
	\caption{Classification accuracy with different numbers of high-pass filters.}
	\label{table-hpf}
	\begin{center}
    \begin{tabular} {|p{3.5cm}|p{1.5cm}|p{1.5cm}|p{1.5cm}|p{1.5cm}|}
    %\begin{tabular}{|l|c|c|c|c|}
    \hline %
     & \multicolumn{2}{|c|}{Image Patches}  & \multicolumn{2}{|c|}{Full-Size Images} \\
    \cline{2-5}
     & model of 50 epochs & model of 80 epochs & model of 50 epochs & model of 80 epochs \\
    \hline \hline
    the proposed HPF$\times$3 & 99.98\% & 99.95\% & 100\% & 100\% \\
    \hline
    the proposed HPF$\times$1 & 99.87\% & 99.77\% & 100\% & 99.83\% \\
    \hline 
    the proposed HPF$\times$0 & 88.28\% & 87.77\% & 93.37\% & 93.12\% \\
    \hline \hline
    Rahmouni~{et al.}~\cite{Rahmouni:2017} & \multicolumn{2}{|c|}{84.8\%} & \multicolumn{2}{|c|}{93.2\%} \\
    \hline %
    \end{tabular}
    \end{center}
\end{table}

\subsubsection{Different Quality Factors of Natural Images}
We also evaluated the robustness of our proposed method with different quality factors. In this experiment, 2000 natural images in RAW format were downloaded from the RAISE-2k dataset~\cite{Dang-Nguyen:2015}. We randomly selected 1800 natural images for our robustness experiment. These RAW images were converted to JPEG format with quality factors of 95, 85, and 75, respectively. Then, we could obtain three sub-datasets with different quality factors of natural images for our experiment. Each of the sub-datasets were then divided into training (50\%), testing (40\%), and validation (10\%) to form the datasets for the robustness experiment of quality factors. Note that the computer-generated graphics in this experiment remained untouched. These computer-generated graphics were compressed with a reasonable quality factor when the author collected this dataset.

\begin{figure}%[htbp]
\centering
\includegraphics[width=12cm]{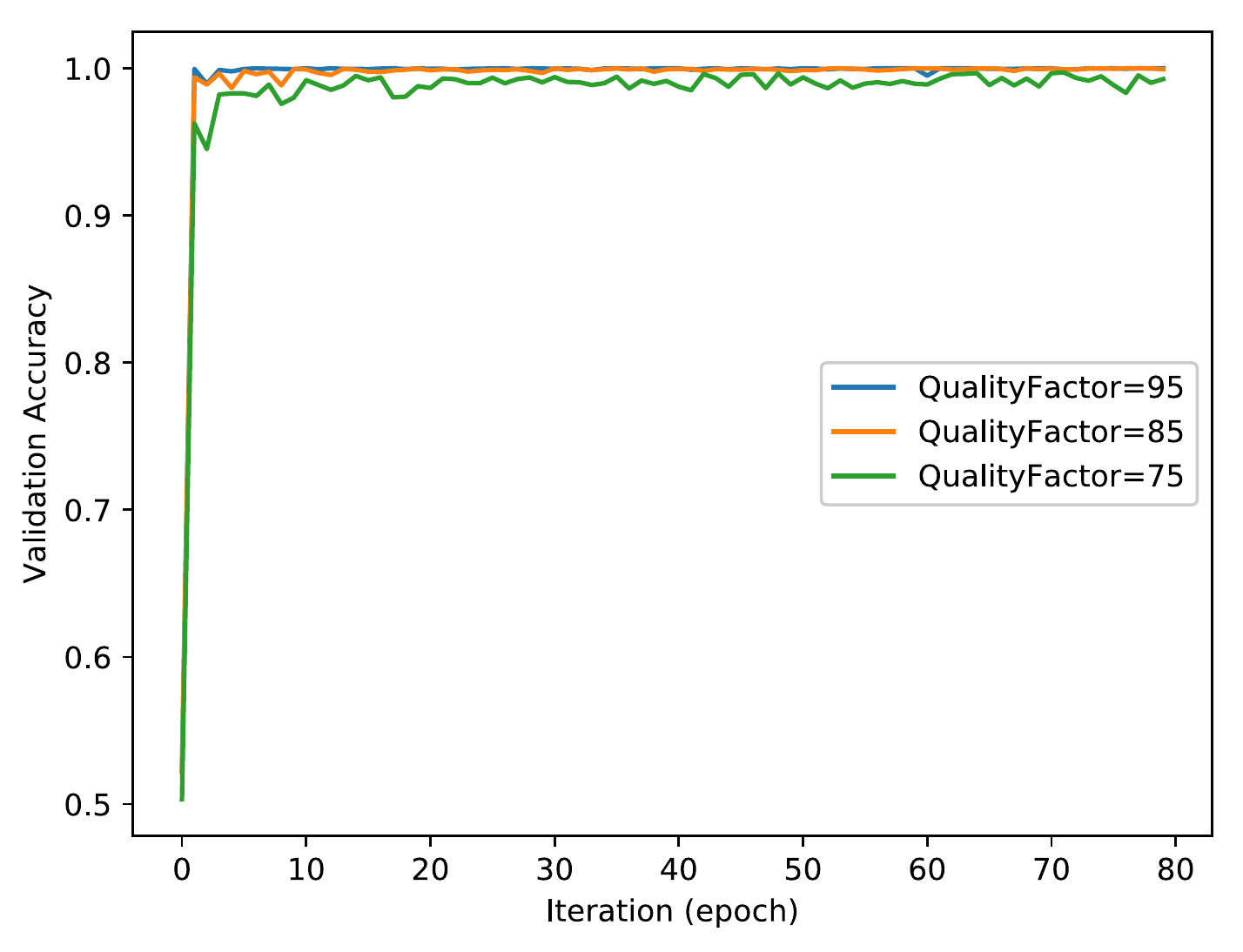}
\caption{Validation performance of the proposed method.}
\label{fig:qf-acc}
\end{figure}

\begin{figure}%[htbp]
\centering
\includegraphics[width=12cm]{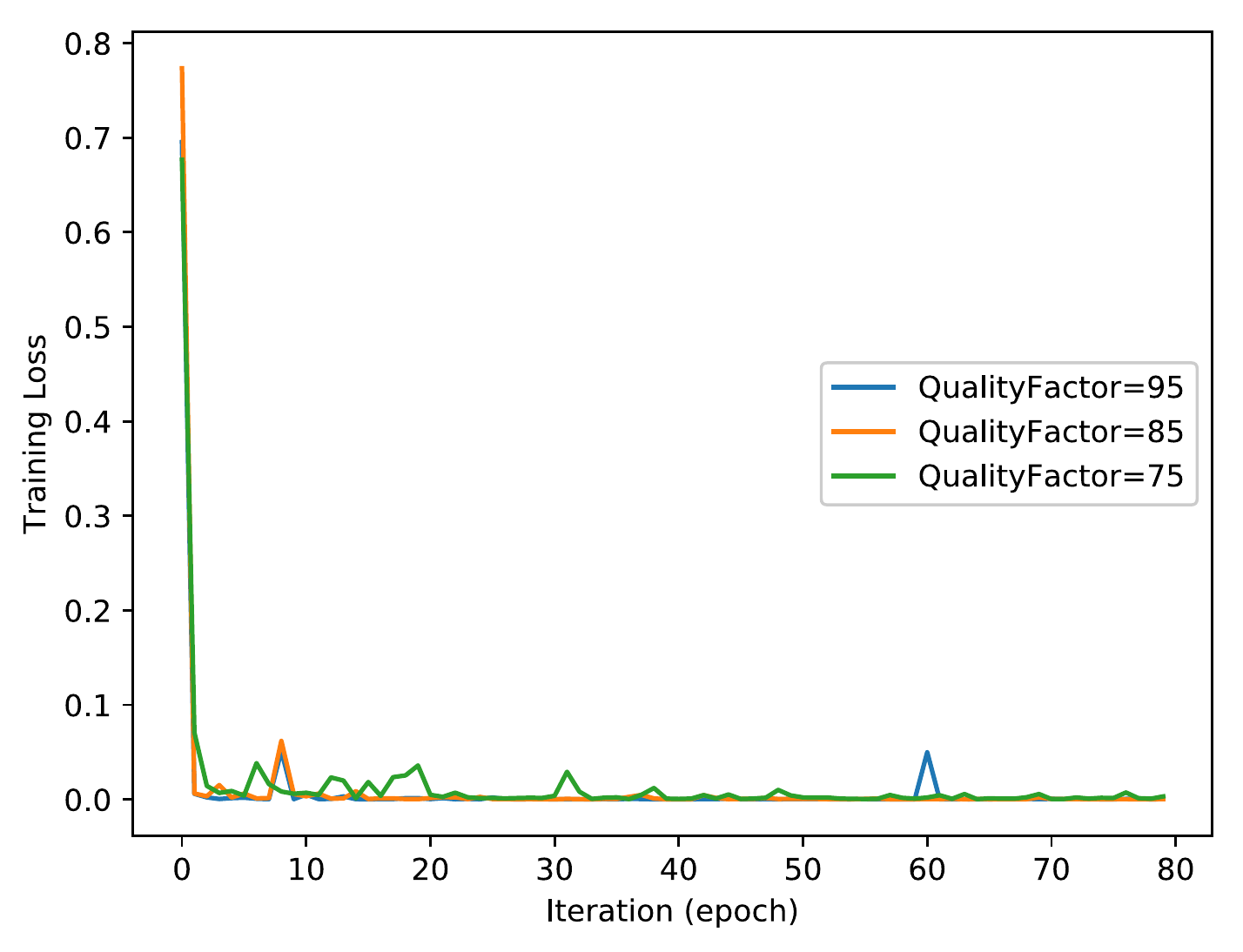}
\caption{Training loss of the proposed method.}
\label{fig:qf-loss}
\end{figure}

For the filter layer, we utilized ~\emph{High\_Pass\_Filter$\times$3} to achieve the best performance in this experiment. Figure~\ref{fig:qf-acc} and Figure~\ref{fig:qf-loss} show the evolutions of training loss and validation accuracy in the iteration procedure. The validation accuracy is shown in Figure~\ref{fig:qf-acc}, and the training loss is shown in Figure~\ref{fig:qf-loss}. The classification accuracy for different quality factors of natural images is shown in Table~\ref{table-qf}. It is observed that the proposed method with~\emph{High\_Pass\_Filter$\times$3} achieves a perfect performance. Although the compression with different quality factors has an impact on the classification accuracy of image patches,  due to the majority vote scheme used for the full-size images, all of the classification accuracies for different quality factors of the natural images are 100\%.

\renewcommand\arraystretch{1.5}
\begin{table}[htbp]
	\caption{Classification accuracy for different quality factors of natural images.}
	\label{table-qf}
	\begin{center}
    \begin{tabular} {|p{3.5cm}|p{1.5cm}|p{1.5cm}|p{1.5cm}|p{1.5cm}|}
    %\begin{tabular}{|l|c|c|c|c|}
    \hline %
     & \multicolumn{2}{|c|}{Image Patches}  & \multicolumn{2}{|c|}{Full-Size Images} \\
    \cline{2-5}
     & model of 50 epochs & model of 80 epochs & model of 50 epochs & model of 80 epochs \\
    \hline \hline
    $QualityFactor=95$ & 99.99\% & 99.99\% & 100\% & 100\% \\
    \hline
    $QualityFactor=85$ & 99.95\% & 99.98\% & 100\% & 100\% \\
    \hline 
    $QualityFactor=75$ & 99.52\% & 99.71\% & 100\% & 100\% \\
    \hline %
    \end{tabular}
    \end{center}
\end{table}

%%%%%%%%%%%%%%%%%%%%%%%%%%%%%%%%%%%%%%%%%%%% section 5
\section{Conclusion}
In this paper, we develop an approach to distinguish computer-generated graphics from natural images based on sensor pattern noise and a convolutional neural network. The experimental results show that the proposed method obtains better performance than the method in~\cite{Rahmouni:2017} on the same dataset. Currently, there are several computer-generated graphics datasets~\cite{Li:2015,Peng:2017} for forensics research. However, many images in these datasets are smaller than 650 pixels in width or height. This cannot meet the size requirement of the proposed convolutional neural network. In the future, we will focus on the improvement of our CNN-based model for smaller images. Furthermore, applying a trained CNN-based model to discriminate the computer-generated graphics from other existing datasets---namely one model for all datasets---would be another interesting future work.

%\newpage
%\IEEEtriggeratref{16}

%\bibliographystyle{IEEEtran}
%\bibliography{mybibs}

\end{document}